# Performance Evaluation of CMOS Annealing with Support Vector Machine

Ryoga Fukuhara
*Graduate School of Informatics*
*Nagoya University*
Aichi, Japan
fukuhara@hpc.itc.nagoya-u.ac.jp

Makoto Morishita
*Graduate School of Informatics*
*Nagoya University*
Aichi, Japan
morishita@hpc.itc.nagoya-u.ac.jp

Takahiro Katagiri
*Information Technology Center*
*Nagoya University*
Aichi, Japan
katagiri@cc.nagoya-u.ac.jp

Masatoshi Kawai
*Information Technology Center*
*Nagoya University*
Aichi, Japan
kawai@cc.nagoya-u.ac.jp

Toru Nagai
*Information Technology Center*
*Nagoya University*
Aichi, Japan
nagai@cc.nagoya-u.ac.jp

Tetsuya Hoshino
*Information Technology Center*
*Nagoya University*
Aichi, Japan
hosino@cc.nagoya-u.ac.jp

***Abstract*—In this paper, support vector machine (SVM) performance was assessed utilizing a quantum-inspired complementary metal-oxide semiconductor (CMOS) annealer. The primary focus during performance evaluation was the accuracy rate in binary classification problems. A comparative analysis was conducted between SVM running on a CPU (classical computation) and executed on a quantum-inspired annealer. The performance outcome was evaluated using a CMOS annealing machine, thereby obtaining an accuracy rate of 93.7% for linearly separable problems, 92.7% for non-linearly separable problem 1, and 97.6% for non-linearly separable problem 2. These results reveal that a CMOS annealing machine can achieve an accuracy rate that closely rivals that of classical computation.**

## I. Introduction

Quantum computers, with their capacity for simultaneous parallel computations, are drawing considerable attention and poised to emerge as the next-generation high-speed computing systems. Owing to their quantum nature, they showcase vastly superior computational capabilities compared to conventional or classical computers. The prediction that Moore's law will eventually plateau, renders substantial speed improvements in classical computers unattainable. Meanwhile, the demand for data processing continues to surge, intensifying the need for high-performance computing solutions, thus elevating the expectations for novel computer paradigms. Consequently, there has been global acceleration in the development of quantum computers.

Diverse quantum annealing methods, semiconductor annealing machines, and other quantum-related hardware exist for addressing combinatorial optimization problems, which are designed to find the optimal combinations of variables that enhance a specific metric among multiple options within various constraints. Notably, semiconductor annealing machines have garnered attention as non-Von Neumann computers that are capable of performing annealing processes to rapidly derive optimal solutions for combinatorial optimization problems at room temperature. However, the range of practical applications of these machines is limited. Therefore, in this study, we implemented a support vector machine (SVM) [1], a well-established machine-learning algorithm, on a CMOS annealing machine, which falls under the category of pseudo-quantum annealers (quantum-inspired annealers). A performance evaluation was conducted on the CMOS annealing machine, which is a type of semiconductor annealing machine, to assess its capabilities.

The SVM is a technique for building a two-class pattern classifier using linear input elements. The parameters of these linear input elements are acquired through training samples with the objective of determining the hyperplane that maximizes the margin distance to each data point. In this process, the hyperparameters governing the misclassification tolerance and generalization performance should be fine-tuned. Consequently, in this study, we focus on optimizing the accuracy rate as an evaluation metric. The aim was to perform a comparative analysis between a CMOS annealing machine and a classical computer to achieve the desired accuracy rate.

The remainder of this paper is organized as follows. Section II offers an introduction to SVM, beginning with a discussion on linearly separable and non-separable problems. Section III provides an explanation of the CMOS annealing machine used in this study. Section IV presents the performance-evaluation results. Section V provides a summary of the relevant research in the field. In the final section, conclusions are drawn.

## II. Overview of Support Vector Machne

### *A. Linearly separable SVM algorithm*

The SVM is a classification method designed to establish decision boundaries that maximize the distance to the nearest data points. In this section, an overview of the algorithm is provided, drawing insights from [2], covering the two cases of: (1) achievable linear separability and (2) non-achievable linear separability.

Linear separability is defined as the capacity to partition a set in n-dimensional space using an n-1-dimensional hyperplane [2] of N point data. The hyperplane equation is defined as

$$\boldsymbol{W}^T\boldsymbol{X} + b = 0, \quad (1)$$

where, $\boldsymbol{X} \in \mathbb{R}^n$, $\boldsymbol{W} \in \mathbb{R}^n$, $b \in \mathbb{R}$. The function returns "1" when the $i$-th data $\boldsymbol{X_i} \in \mathbb{R}^n$ belongs to class 1. and returns "-1" when it belongs to class 2, Introducing $t_i \in \mathbb{R}$, the condition equation is:

$$t_i(\boldsymbol{W}^T\boldsymbol{X}_i + b) > 0 \ (i = 1,2,3, \dots N) \quad (2)$$

The distance $d$ between point $\boldsymbol{X}_i$ on the *n*-dimensional space and the hyperplane is described as:

$$d = \frac{|w_1x_{1_i} + w_2x_{2_i} + \cdots + w_nx_{n_i} + b|}{\sqrt{w_1^2 + w_2^2 + \cdots + w_n^2}} = \frac{|\boldsymbol{W}^T\boldsymbol{X}_i + b|}{\|\boldsymbol{W}\|} \quad (3)$$

Using Eqs. (1)-(3), the condition for maximizing margin *M* is

$$max_{w,b}M, \ \frac{t_i(\boldsymbol{W}^T\boldsymbol{X}_i + b)}{\|\boldsymbol{W}\|} \geq M \ (i = 1,2,3, \dots N) \quad (4)$$

Normalizing with $\|W\| = \frac{1}{M}$, Eq. (5) is obtained.

$$max_{w,b}\frac{1}{\|\boldsymbol{W}\|}, \ t_i(\boldsymbol{W}^T\boldsymbol{X}_i + b) \geq 1 \ (i = 1,2,3, \dots N) \quad (5)$$

To minimize the energy function, Eq. (6) can be derived as

$$min_{w,b}\frac{1}{2}\|\boldsymbol{W}\|^2, \quad t_i(\boldsymbol{W}^T\boldsymbol{X}_i + b) - 1 \geq 0 \quad (i = 1,2,3, \dots N) \ (6)$$

*B. Linearly non-separable SVM algorithm*

When linear separation is unattainable, primarily two approaches are considered: (1) tolerating misclassification, and (2) transforming into high-dimensional spaces [2]. Employing both methods simultaneously is advisable.

If linear separability is not possible, then the constraint in Eq. (6) is not satisfied and learning is not possible. Thus, slack variable $\varepsilon_i \in \mathbb{R}$, expressed in Eq. (7), was introduced.

$$\varepsilon_i = \max\{0, 1 - t_i(\boldsymbol{W}^T\boldsymbol{X}_i + b)\}, \quad (7)$$

where $t_i \in R$.

In this context, the slack variable $\varepsilon_i$ $is\ introduced$ into the constraint condition, permitting a certain level of misclassification. Consequently, we obtain Eq. (8).

$$t_i(\boldsymbol{W}^T\boldsymbol{X}_i + b) - 1 + \varepsilon_i \geq 0 \ (i = 1,2,3, \dots N) \quad (8)$$

Properties of the slack variable $\varepsilon_i$ can be summarized, as follows:

- If $0 \leq \varepsilon_i \leq 1$, it falls within the margin range.
- Misclassification occurs when $1 < \varepsilon_i$.
- The degree of misclassification increases for larger $\varepsilon_i$.

Next, Eq, (8) is modified into a function that allows misclassification, thereby obtaining Eq. (9).

$$min_{w,b}\frac{1}{2}\|W\|^2 + C\sum_{i=1}^{n}\varepsilon_i,$$

$$t_i(W^TX_i + b) - 1 + \varepsilon_i \geq 0 \ (i = 1,2,3, \dots N) \ (9)$$

The larger the coefficient $C$ in Eq. (9), the greater is its influence on the misclassification minimization function Therefore, C has the following properties:

- Large $C$: can lead to overfitting.
- Small $C$: can lead to generalization and not learning.

A solution is obtained by transforming a linearly non-separable problem into linearly separable high-dimensional coordinates through the mapping $\varphi$ and subsequently reversing the transformation. However, due to the high computational cost associated with calculating the inner product post-projection, a direct definition of mapping $\varphi$ is not feasible.

In this study, the radial basis function (RBF) kernel [3], described by Eq. (10) is used.

$$K(X_i, X_j) = \exp\left(-\frac{\|X_i - X_j\|^2}{2\sigma^2}\right) = \exp(-\gamma\|X_i - X_j\|^2), \quad (10)$$

where, the parameter $\gamma \in \mathbb{R}$ represents "the range within which an individual data point influences the decision boundary." When $\gamma$ is increased, the influence range of an individual data point decreases, resulting in a decision boundary with a more pronounced curvature.

## III. CMOS ANNERING MACHINE

*A. Ising model*

The Ising model [5] is a statistical mechanical model employed to characterize the behavior of magnetic materials, including magnets. This model exhibits the following key characteristics:

- It comprises spins represented as lattice points that can exist in one of two states: upward or downward.
- The states of neighboring spins are subject to updates influenced by interactions under the influence of an externally applied magnetic field.

- Ultimately, the spins reach a stable state when the energy of the Ising model reaches its minimum value.

This model can be mathematically formulated as follows:

$$H = \sum_{i \neq j} J_{ij}\sigma_i\sigma_j + \sum_i h_i\sigma_i \quad (\sigma = \pm 1) \tag{11}$$

In Eq. (11), $\sigma_i$ represents the input variable and is commonly referred to as a spin. In the context of quantum annealing, the parameters $J_{ij}$, associated with qubits, represent interaction parameters, whereas $h_i$ is a single parameter referred to as the magnetic field.

The Ising model is highly versatile and serves as an effective means for expressing combinatorial optimization problems in a general and adaptable manner. as well as a suitable mathematical model for quantum-annealing machines. Consequently, the Ising model is frequently employed as the input to annealing machines, owing to its compatibility and versatility in handling a wide range of optimization problems.

## IV. Performance Evaluation

### A. *Problem setting*

In performance evaluation, we addressed the task of generating two-dimensional random numbers, categorizing them into two classes based on whether they exceeded or fell below a specified functional threshold. The set of functions was altered to create three distinct problem types: one that was linearly separable and two that were not.

To address this problem, 100 data points were utilized as the training data and 1000 data points as the test data. In addition, training data with a 5% misclassification rate were generated by randomly selecting points. These datasets are displayed in Figs. 1 - 3.

In this study, the solution was assessed in terms of accuracy, which refers to the percentage of correctly provided predefined classification outcomes.

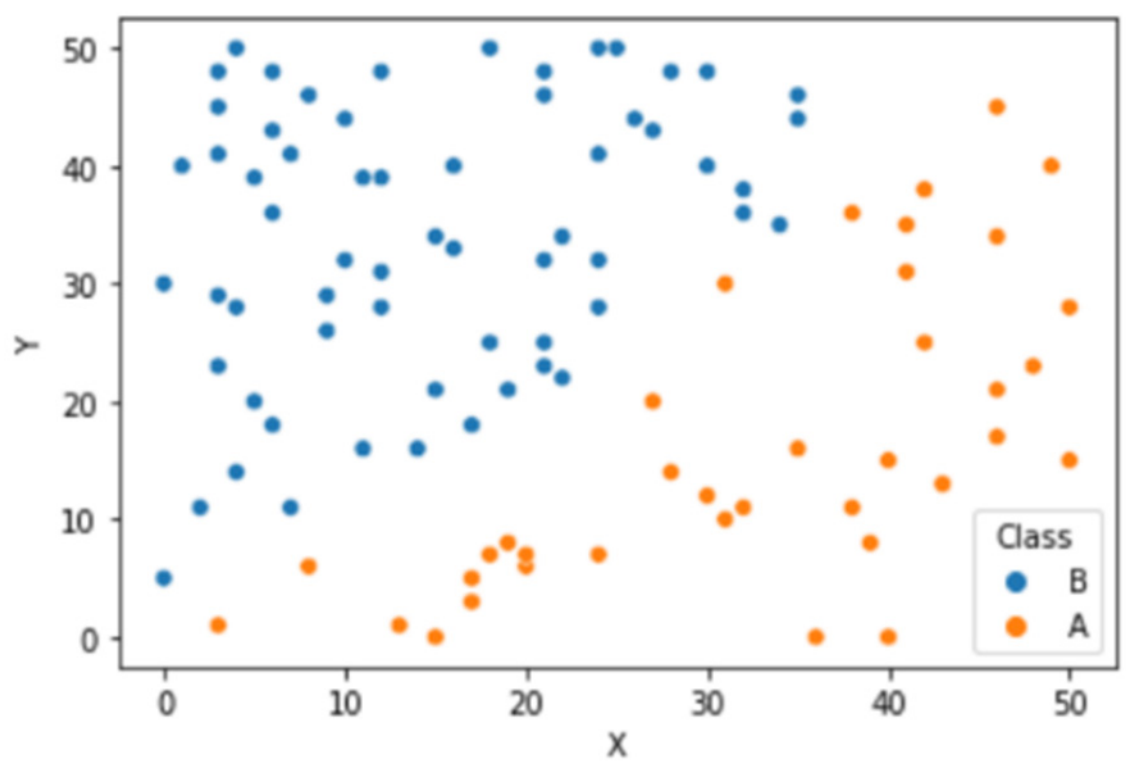

(a) Learning data without error.

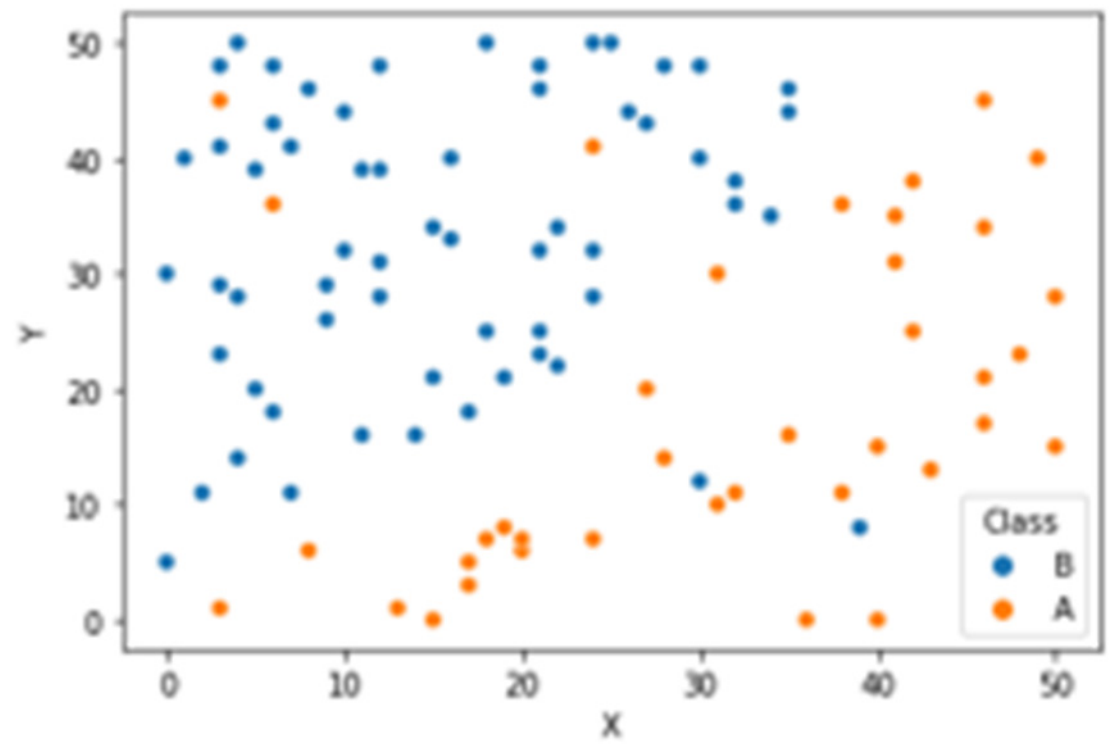

(b) Learning data with 5% error

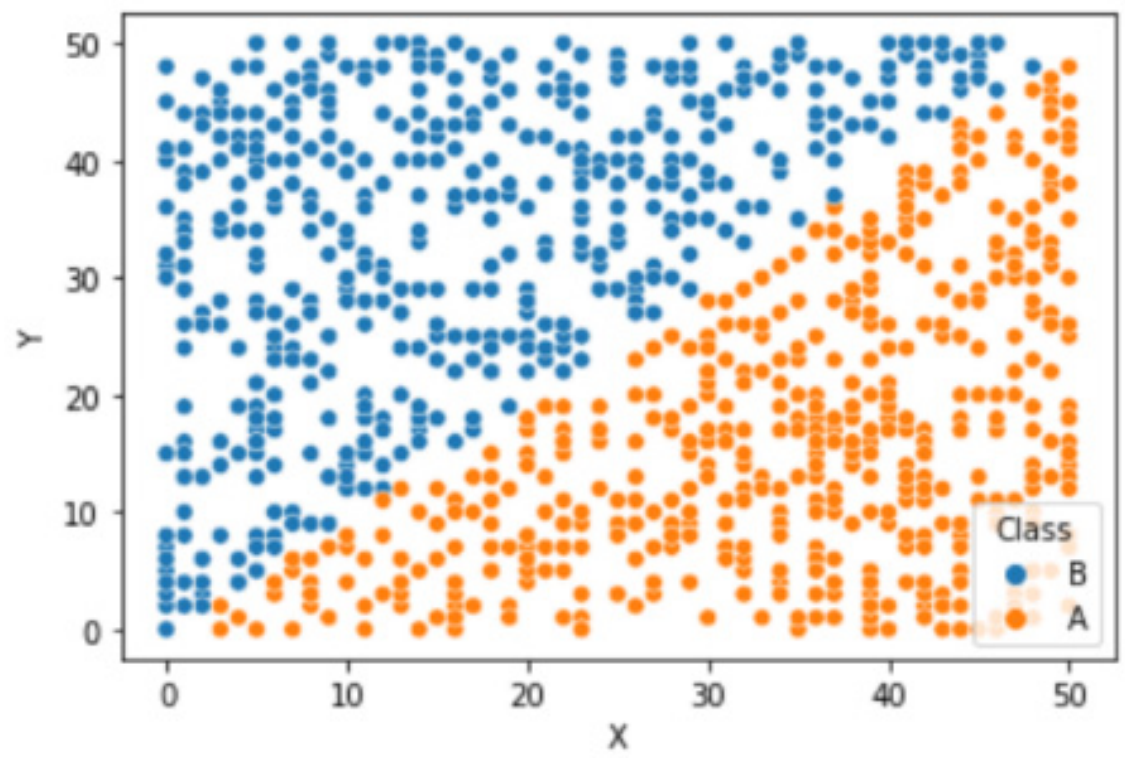

(c) Test data (true labeled data).

Fig. 1. Dataset for linearly separable case.

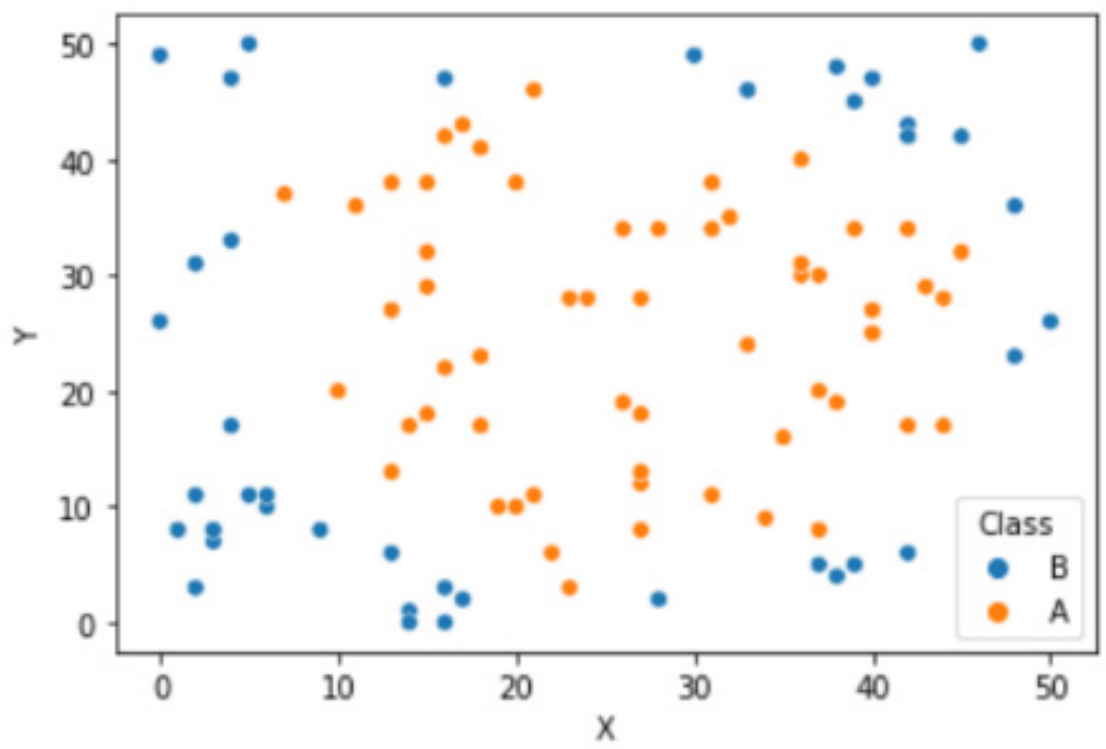

(a) Learning data without error.

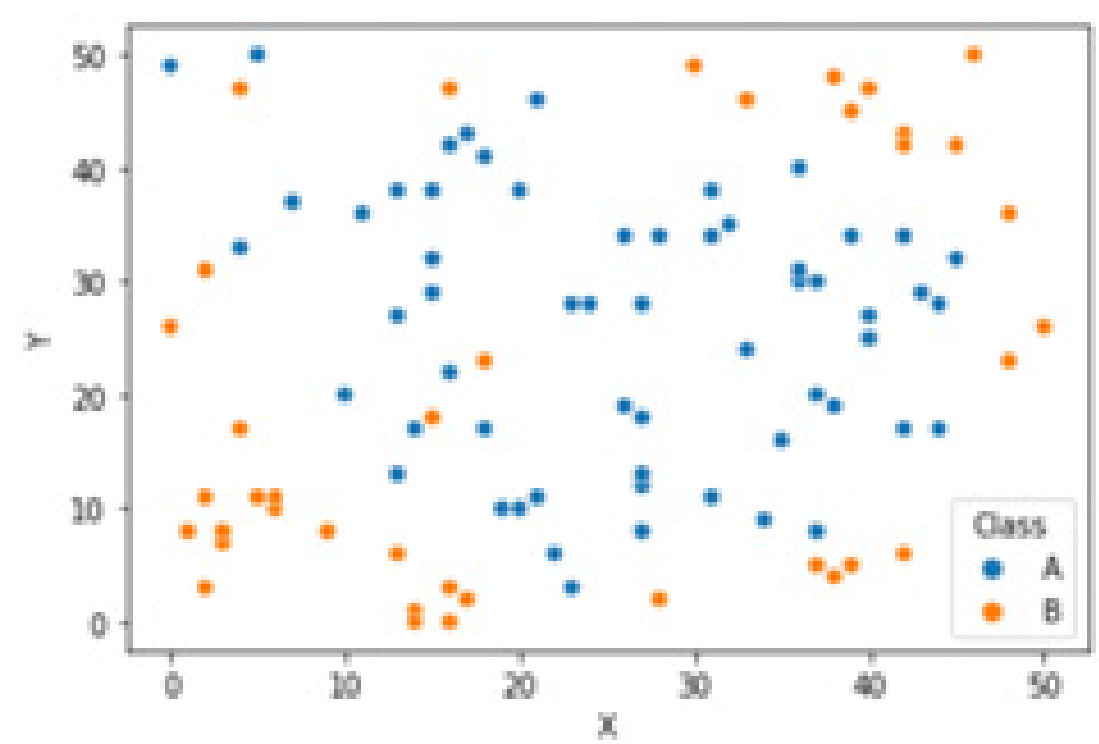

(b) Learning data with 5% error

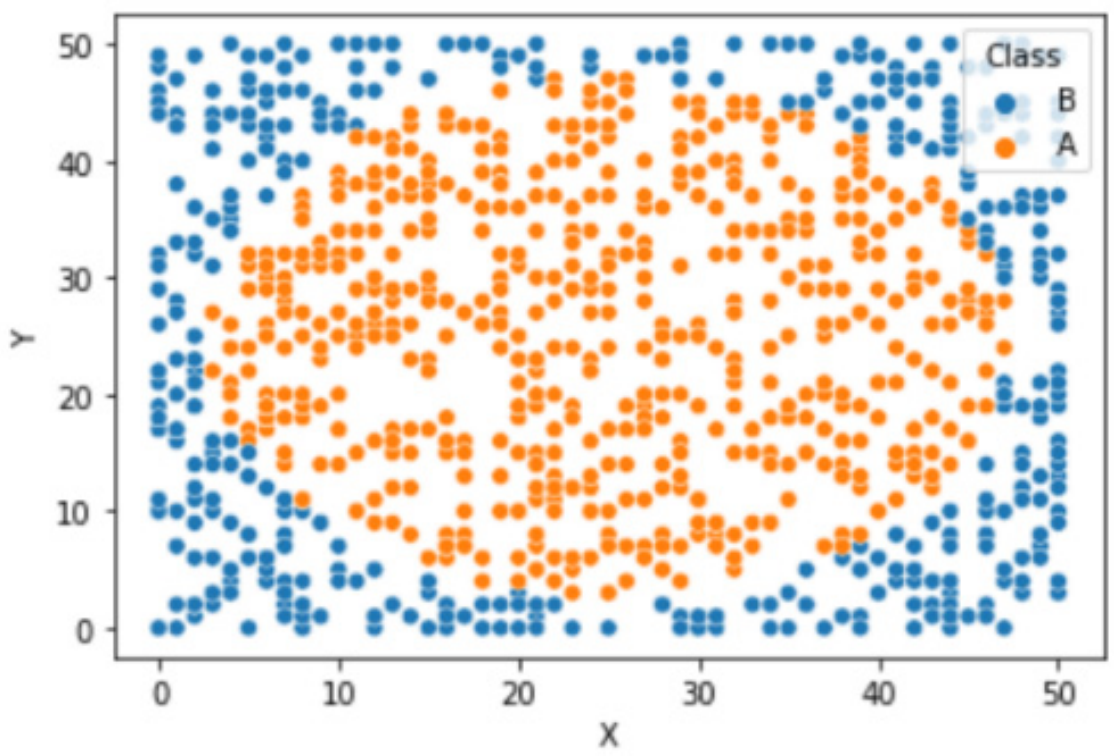

(c) Test data (true labeled data).

Fig. 2. Dataset for non-linearly separable case 1.

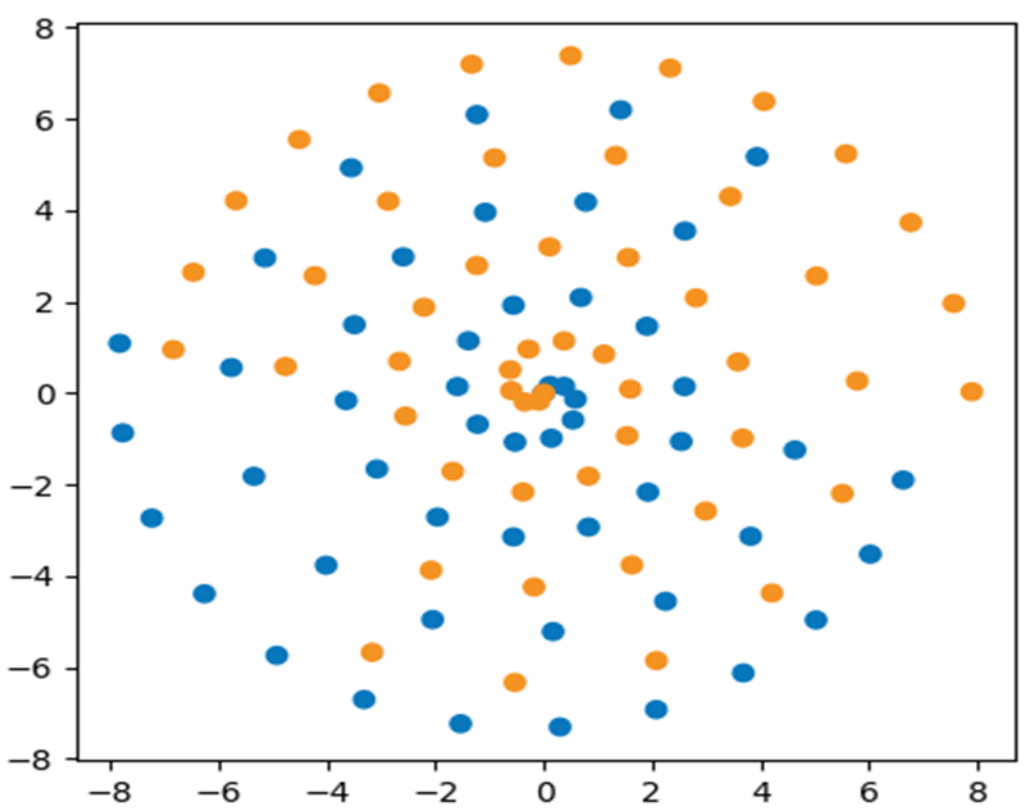

(a) Learning data without error.

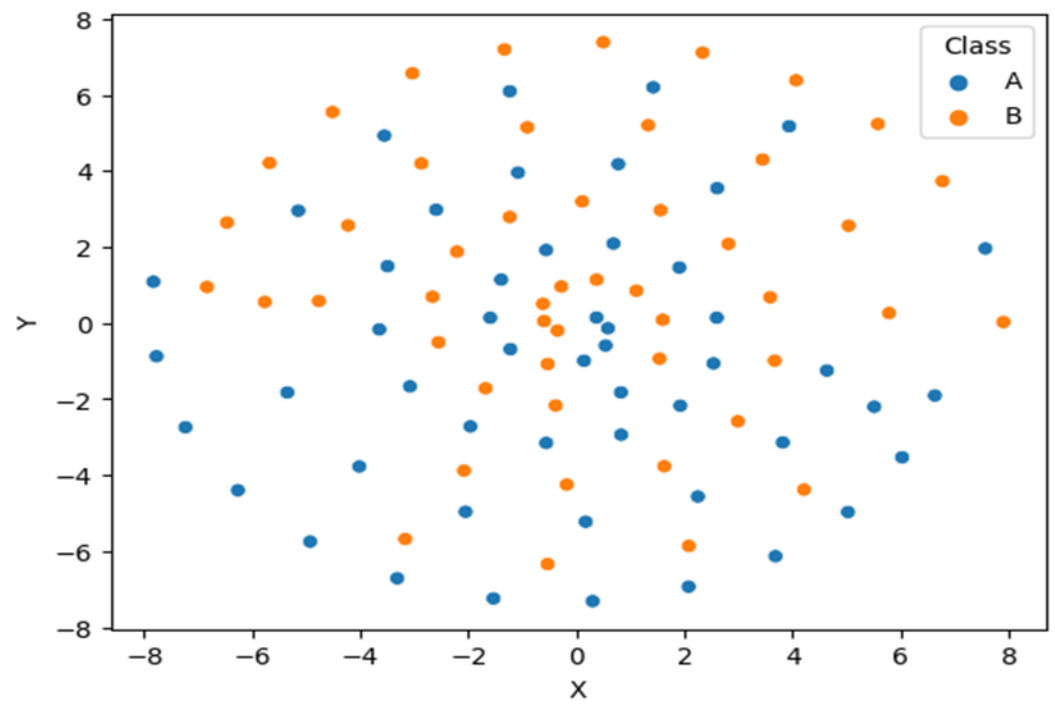

(b) Learning data with 5% error

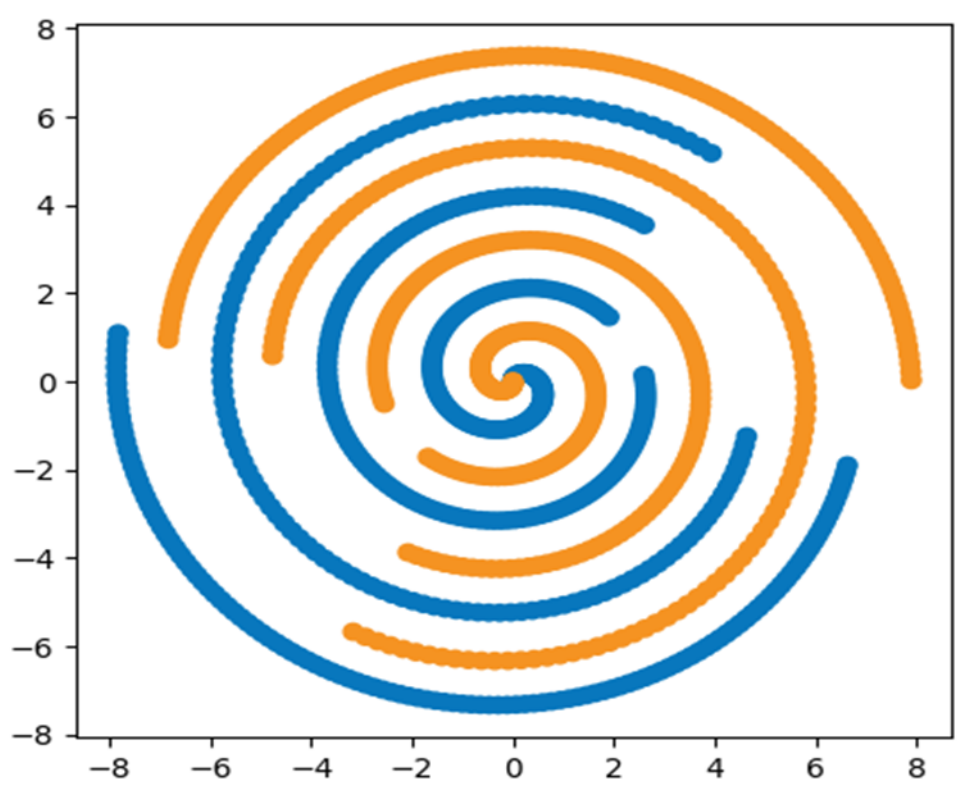

(c) Test data (labeled data).

Fig. 3. Dataset for non-linearly separable case (2).

## B. Machine environment

Google Colaboratory was used as the computational environment for CPU processing. The CMOS annealing machine utilized in this study was Fixstars' Amplify [7], corresponding to the equipment employed in [6].

The CMOS annealing environment was Annealing Cloud Web API version 2: GPU version (32-bit float) and Amplify version 0.5.13. The classical environment consisted of Intel Core i7-8550U CPU (1.80 GHz, up to 2.00 GHz) and Google Colaboratory Python version 3.7.12.

## C. Procedure

The procedure consisted of the following steps:

1. Create an SVM classifier from training data with 100 samples.
2. Apply the obtained classifier to the 1000 samples of the test data (true labeled data) and calculate the confusion matrix.
3. Calculate the accuracy rate from the confusion matrix.

Steps 1–3 were performed on a conventional computer and CMOS annealing machine for the linearly separable problem, thereby evaluating the performance.

For this evaluation, the Hamiltonians in Eq. (12) was used for CMOS annealing [8].

$$H = \frac{1}{2}\sum_{n,m,k,j} a_{K_{n+k}} a_{K_{m+j}} B^{k+j} t_n t_m k(x_n, x_m) - \sum_{n,k} B^k a_{K_{n+k}} + \frac{1}{2}\xi\Big(\sum_{n,k} B^k a_{K_{n+k}} t_n\Big)^2 \quad (12)$$

Table I shows four hyperparameters: $B, K, \gamma$ and $\xi$. The accuracy of each parameter was examined within the ranges listed in Table I. Subsequently, the parameters with the highest accuracy were adopted.

TABLE I. HYPER PARAMETERS ON CMOS ANNEALING MACHINE

| Hyper Parameters | Values |
|---|---|
| Base of encoding $B$ | 2, 10 |
| Number of binary variables for encoding number $K$ | 2, 3 |
| Coefficient in RBF kernel function $\gamma$ | 0.0001, 0.001, 0.01, 0.1, 1, 10, 100, 1000 |
| Coefficient of second constraint in the energy function $\xi$ | 0, 10, 100 |
| Number of iterations for annealing | 1 |

Labeling the data corresponding to the correct answers, the accuracy was calculated from Steps 1-3 using Eq. (13).

$$Precidion = \frac{Number\ of\ correct\ labels\ predicted}{1000} \quad (13)$$

## D. Results

Table II displays the maximum and minimum accuracies for each problem.

TABLE II. MAXIMUM AND MINIMAL ACCURACIES FOR EACH PROBLEM FOR CMOS ANNEALING

| Problem Kinds | | Linearly Separable | Linearly Non-Separable (1) | Linearly Non-Separable (2) |
|---|---|---|---|---|
| *CMOS annealing without error* | Maximal Accuracy [%] | 93. 7% | 92.7% | **97.6%** |
| | Minimal Accuracy [%] | 49.6% | 49.4% | 50.6% |
| | Diff. [Point] | 44.1 | 43.3 | 47.0 |
| *CMOS annealing with error 5%* | Maximal Accuracy [%] | 93.5% | 91.1% | 92.5% |
| | Minimal Accuracy [%] | 49.6% | 44.0% | **40.2%** |
| | Diff. [Point] | 43.9 | 47.1 | **52.3** |

The impact of hyperparameter tuning is significant in Table I, with the most substantial variance in accuracy observed in the linearly non-separable case 2, where the difference is 52.3 points. In this scenario, the minimum and maximum accuracies are 40.2% and 92.5%, respectively. This substantial difference underscores the importance of hyperparameter tuning, shown in Table I, as a crucial step toward achieving highly accurate solutions in a quantum-inspired annealer.

Table III presents the accuracy of each problem evaluated using both classical and CMOS annealers.

According to the results in Table III, the CMOS annealer consistently achieves an accuracy that either equals or surpasses that of classical computers. In this context, the solution accuracy does not appear to be a concern when using the CMOS annealer, compared to the classical approach.

However, in the case of nonlinear separability (2), the CMOS annealer outperforms the classical solution in terms of accuracy. The exact cause of this discrepancy remains unknown, but potential factors may include (1) disparities in the expressions used for SVM between the two approaches and (2) the possibility that parameter tuning in the classical approach depends on the tool used, making manual parameter tuning a potential consideration.

TABLE III. SUMMARY OF ACCURACIES

| Problem Kinds | | Linearly Separatable | Linearly Non-Separable (1) | Linearly Non-Separable (2) |
|---|---|---|---|---|
| *Classical without error* | Accuracy [%] | 93.0% | 85.8% | 63.3% |
| *CMOS annealing without error* | Accuracy [%] | **93.7%** | **92.7%** | **97.6%** |
| | Specify Parameters $(B, K, \gamma,\ \xi)$ | (10, 3, 0.001, 100) | (10, 2, 1, 0) | (2, 2, 100, 0) |
| *Classical with error 5%* | Accuracy [%] | 85.7% | 85.4% | 66.7% |
| *CMOS annealing with error 5%* | Accuracy [%] | **93.5%** | **91.1%** | **92.5%** |
| | Specify Parameters $(B, K, \gamma,\ \xi)$ | (10, 2, 0.1, 100) | (10, 2, 1, 0) | (2, 3, 10, 100) |

## E. Dissussion

From the data presented in Table III, CMOS annealing achieves an accuracy rate that is either close to or higher than that of classical computers. The research in [9] achieved an accuracy rate of approximately 85.8% when dealing with similar linearly separable problems. The enhanced accuracy observed in this experiment can be attributed to two key factors.

1. Increased search range for hyperparameters: the search range for hyperparameters was expanded compared to that in [9], which possibly contributed to improved performance.

2. Larger batch size: Another significant change was increasing the number of data points used for training from 100 to 1000. This adjustment in batch size can substantially impact learning and generalization.

Table II illustrates the substantial influence of hyperparameter optimization (point 1) on the accuracy of the solution, which ranges from 40.0% to 97.6% as hyperparameters are adjusted. This underscores the importance of fine-tuning the parameters embedded in the Hamiltonian used for SVM in the CMOS annealer, as these adjustments play a crucial role in enhancing the accuracy of the solution.

Furthermore, CMOS demonstrated higher accuracy than classical methods, primarily because of the inadequate tuning of the hyperparameters in the classical approach. Hyperparameter tuning for classical SVM was omitted in this experiment. Implementing hyperparameter tuning for classical SVM would mitigate the accuracy gap between classical SVMs and CMOS.

In addition, the start-up time of CMOS annealing is approximately 1000 times longer than that of the classical method. This extended duration is partly attributed to suboptimal Python calculations, making it imperative to optimize the CMOS annealing process.

## V. Related Work

In this study, a quantum-inspired annealer was used for the experiments. However, the most intriguing point in the comparisons was the capability of the true quantum annealer to harness quantum properties. The study in [8] applied SVM using a quantum annealer, whereas the study in [7] involved an experiment with the D-wave machine, which is an actual quantum annealer. In contrast, our research employed Hitachi's CMOS annealing, which does not fully leverage quantum characteristics. Additionally, our study differs in terms of using a diverse range of data unlike the two-dimensional synthetic data used in [8].

Notably, the argument in [8] suggests that quantum annealers are effective, particularly when dealing with limited training data. This perspective makes it imperative to evaluate whether the efficiency demonstrated by the pseudo-quantum annealer, driven by high speed and improved solution accuracy, can compete with that of quantum annealers.

In this study, we demonstrated that the accuracy of the SVM solution exhibits substantial variability based on the fine-tuning of the Hamiltonian hyperparameters required for quadratic unconstrained binary optimization (QUBO). In other words, these hyperparameter adjustments are pivotal factors that influence the accuracy of the SVM solution in a quantum-inspired annealer. Furthermore, this research underscores the significance of weight adjustments for the constraint terms in QUBO through performance evaluations extending beyond SVM [8].

The adjustments of SVM-specific hyperparameters within the quantum annealer and hyperparameters for the constraint and optimization terms within QUBO are of paramount importance in practical applications. The challenge of fine-tuning performance-related hyperparameters aligns with the problems of software automatic tuning (AT) [10]. Therefore, applying AT techniques to optimize performance parameters in SVM, proposing SVM-specific AT methods, and assessing performance using AT technology are important venues to explore in future work.

## VI. Conclusion

In this study, support vector machines were evaluated in a pseudo-quantum annealing environment. Using the CMOS annealer developed by Hitachi as a quantum-inspired annealer, we formulated artificial problems encompassing two categories of binary classification problems: linearly separable and linearly non-separable, thereby evaluating the performance of SVM within this context.

In the evaluations conducted using Amplify, a quantum-inspired annealer environment in the cloud, the following accuracy rates were obtained: (i) linearly separable problems: 93.7%; (ii) linearly non-separable problems (1): 92.7%; and (iii) linearly non-separable problems (2): 97.6%.

Furthermore, in performance evaluations in a classical computing environment, the accuracy rates matched those achieved in the pseudo-quantum annealing environment.

However, the evaluations also highlighted the crucial role of adjusting SVM-specific hyperparameters during execution within a quantum-inspired annealer, which had a significant impact on the solution accuracy. Consequently, automating this hyperparameter adjustment process is imperative given that manual adjustment incurs human and time costs.

The optimization of performance parameters for both SVM-specific quantum-inspired annealers and quantum annealers can be enhanced by utilizing auto-tuning (AT) technology. With a proven track record in this field, we adapted AT to quantum-circuit simulations, as documented in [11]. The integration of AT into quantum-related technology is gaining momentum. The application and subsequent evaluation of AT techniques represent crucial areas for future research with potential to streamline and enhance the efficiency of both quantum and quantum-inspired annealers.

## Acknowledgment

This work was supported by JSPS KAKENHI Grant Number JP19H05662, and by “Joint Usage/Research Center for Interdisciplinary Large-scale Information Infrastructures (JHPCN)” in Japan (Project ID: jh230005 and jh230053).

We would like to thank Mr. Kazuo Ono and Mr. Masanao Yamaoka of Hitachi, Ltd. for their advice regarding the use of CMOS Annealer.

We would like to thank Yoshiki Matsuda of Fixstars Corporation for advice regarding the use of Amplify.

## References

[1] V. N. Vapnik, “The Nature of Statistical Learning Theory.”, Springer, New York, 1995.

[2] D. P. Kroese, et al, “Data Science and Machine Learning -From Theory to Implementation with Python-”, Tokyo Kagaku Dojin, ISBN :9784807920297, 2022, pp.212. In Japanese.

[3] D. P. Kroese, et al, “Data Science and Machine Learning -From Theory to Implementation with Python-”, Tokyo Kagaku Dojin, ISBN :9784807920297, 2022, p.176, 2022. In Japanese.

[4] M. Yamaoka, C. Yoshimura, M. Hayashi, T. Okuyama, H. Aoki, H. Mizuno, “A 20k-spin Ising chip for combinatorial optimization problems with CMOS annealing”, In Proc. of IEEE International Solid-State Circuits Conference, Vol. 51, No. 1, pp. 303 – 309, January 2016. https://doi.org/10.1109/JSSC.2015.2498601

[5] CMOS Annealing Machine - Annealing Cloud Web. https://annealing-cloud.com

[6] M. Morishita, T. Katagiri, S. Ohshima, T. Nagai, “Analysis of Performance for CMOS annealing machine with Amplify”, IPSJ SIG Technical Report, Vol. 2021-HPC-181, No. 3, pp. 1-6, September 20, 2021. In Japanese.

[7] Fixters Amplify. https://amplify.fixstars.com/ja/

[8] D. Willsch, M. Willsch, J. De Raedt, K. Michielsen, "Support Vector Machines on the D-Wave Quantum Annealer", Computer Physics Communications, Vol. 248, pp.107006, March 2020. https://doi.org/10.1016/j.cpc.2019.107006

[9] R. Fukuhara, M. Morishita, T. Katagiri, S. Ohshima, T. Nagai, "Investigation of Applicability of Combinatorial Optimization Problems in Quantum Annealing Machines", In Proc. of The 84th National Convention of IPSJ, Vol. 2, pp. 69-70, March 2022. In Japanese.

[10] T. Katagiri, D. Takahashi, "Japanese Auto-tuning Research: Auto-tuning Languages and FFT", In Proc. of the IEEE, Vol. 106, No. 11, pp. 2056 - 2067, November 2018. https://doi.org/10.1109/JPROC.2018.2870284

[11] M. Morishita, T. Katagiri, S. Ohshima, T. Hoshino, T. Nagai, "Application of Auto-tuning to Quantum Computing", IPSJ SIG Technical Report, Vol. 2023-HPC-188, No. 2, pp.1-7, March 2023. In Japanese.

## APPENDIX

Table AI shows the parameter tuning history of linearly non-separatable (2).

Table A1: Accuracies for varying hyper parameters on linearly non-separatable (2)

| $B$ | $K$ | $\gamma$ | $\xi$ | Accuracy |
|---|---|---|---|---|
| 2 | 2 | 0.0001 | 0 | 0.621 |
| 2 | 2 | 0.0001 | 10 | 0.622 |
| 2 | 2 | 0.0001 | 100 | 0.622 |
| 2 | 2 | 0.001 | 0 | 0.673 |
| 2 | 2 | 0.001 | 10 | 0.673 |
| 2 | 2 | 0.001 | 100 | 0.673 |
| 2 | 2 | 0.01 | 0 | 0.659 |
| 2 | 2 | 0.01 | 10 | 0.659 |
| 2 | 2 | 0.01 | 100 | 0.659 |
| 2 | 2 | 0.1 | 0 | 0.655 |
| 2 | 2 | 0.1 | 10 | 0.655 |
| 2 | 2 | 0.1 | 100 | 0.655 |
| 2 | 2 | 1 | 0 | 0.658 |
| 2 | 2 | 1 | 10 | 0.658 |
| 2 | 2 | 1 | 100 | 0.658 |
| 2 | 2 | 10 | 0 | 0.891 |
| 2 | 2 | 10 | 10 | 0.887 |
| 2 | 2 | 10 | 100 | 0.887 |
| **2** | **2** | **100** | **0** | **0.925** |
| 2 | 2 | 100 | 10 | 0.877 |
| 2 | 2 | 100 | 100 | 0.877 |
| 2 | 2 | 1000 | 0 | 0.764 |
| 2 | 2 | 1000 | 10 | 0.635 |
| 2 | 2 | 1000 | 100 | 0.633 |
| 2 | 3 | 0.0001 | 0 | 0.66 |
| 2 | 3 | 0.0001 | 10 | 0.648 |
| 2 | 3 | 0.0001 | 100 | 0.656 |
| 2 | 3 | 0.001 | 0 | 0.664 |
| 2 | 3 | 0.001 | 10 | 0.662 |
| 2 | 3 | 0.001 | 100 | 0.663 |
| 2 | 3 | 0.01 | 0 | 0.656 |
| 2 | 3 | 0.01 | 10 | 0.656 |
| 2 | 3 | 0.01 | 100 | 0.653 |
| 2 | 3 | 0.1 | 0 | 0.65 |
| 2 | 3 | 0.1 | 10 | 0.644 |
| 2 | 3 | 0.1 | 100 | 0.627 |
| 2 | 3 | 1 | 0 | 0.665 |
| 2 | 3 | 1 | 10 | 0.659 |
| 2 | 3 | 1 | 100 | 0.608 |
| 2 | 3 | 10 | 0 | 0.904 |
| 2 | 3 | 10 | 10 | 0.891 |
| 2 | 3 | 10 | 100 | 0.859 |
| 2 | 3 | 100 | 0 | 0.925 |
| 2 | 3 | 100 | 10 | 0.831 |
| 2 | 3 | 100 | 100 | 0.82 |
| 2 | 3 | 1000 | 0 | 0.764 |
| 2 | 3 | 1000 | 10 | 0.592 |
| 2 | 3 | 1000 | 100 | 0.601 |
| 10 | 2 | 0.0001 | 0 | 0.674 |
| 10 | 2 | 0.0001 | 10 | 0.664 |
| 10 | 2 | 0.0001 | 100 | 0.657 |
| 10 | 2 | 0.001 | 0 | 0.664 |
| 10 | 2 | 0.001 | 10 | 0.643 |
| 10 | 2 | 0.001 | 100 | 0.594 |
| 10 | 2 | 0.01 | 0 | 0.659 |
| 10 | 2 | 0.01 | 10 | 0.654 |
| 10 | 2 | 0.01 | 100 | 0.571 |
| 10 | 2 | 0.1 | 0 | 0.663 |
| 10 | 2 | 0.1 | 10 | 0.615 |
| 10 | 2 | 0.1 | 100 | 0.522 |
| 10 | 2 | 1 | 0 | 0.623 |
| 10 | 2 | 1 | 10 | 0.585 |
| 10 | 2 | 1 | 100 | 0.559 |
| 10 | 2 | 10 | 0 | 0.834 |
| 10 | 2 | 10 | 10 | 0.645 |
| 10 | 2 | 10 | 100 | 0.576 |
| 10 | 2 | 100 | 0 | 0.925 |
| 10 | 2 | 100 | 10 | 0.818 |
| 10 | 2 | 100 | 100 | 0.812 |
| 10 | 2 | 1000 | 0 | 0.764 |
| 10 | 2 | 1000 | 10 | 0.686 |
| 10 | 2 | 1000 | 100 | 0.547 |
| 10 | 3 | 0.0001 | 0 | 0.655 |
| 10 | 3 | 0.0001 | 10 | 0.635 |
| 10 | 3 | 0.0001 | 100 | 0.654 |
| 10 | 3 | 0.001 | 0 | 0.68 |
| 10 | 3 | 0.001 | 10 | 0.655 |
| 10 | 3 | 0.001 | 100 | 0.641 |
| 10 | 3 | 0.01 | 0 | 0.656 |
| 10 | 3 | 0.01 | 10 | 0.574 |
| **10** | **3** | **0.01** | **100** | **0.402** |
| 10 | 3 | 0.1 | 0 | 0.591 |
| 10 | 3 | 0.1 | 10 | 0.522 |
| 10 | 3 | 0.1 | 100 | 0.467 |
| 10 | 3 | 1 | 0 | 0.499 |
| 10 | 3 | 1 | 10 | 0.519 |
| 10 | 3 | 1 | 100 | 0.528 |
| 10 | 3 | 10 | 0 | 0.834 |
| 10 | 3 | 10 | 10 | 0.576 |
| 10 | 3 | 10 | 100 | 0.652 |
| 10 | 3 | 100 | 0 | 0.925 |
| 10 | 3 | 100 | 10 | 0.543 |
| 10 | 3 | 100 | 100 | 0.775 |
| 10 | 3 | 1000 | 0 | 0.764 |
| 10 | 3 | 1000 | 10 | 0.682 |
| 10 | 3 | 1000 | 100 | 0.549 |